\documentclass[twocolumn,showpacs,preprintnumbers,amsmath,amssymb,aps,prb]{revtex4}

\usepackage{graphicx}
\usepackage{dcolumn}
\usepackage{bm}

\begin{document}

\title{Conductance distributions of 1D-disordered wires at finite temperature 
and bias voltage.}

\author{Federico Foieri}
\affiliation{Departamento de F{\'i}sica ``J. J. Giambiagi" FCEyN, Universidad 
de Buenos Aires, 
Ciudad Universitaria Pab.I, (1428) Buenos Aires, Argentina}

\author{Mar\'{\i}a Jos\'e  S\'anchez}
\affiliation{Centro At\'omico Bariloche and Instituto Balseiro, Bustillo 
9500 (8400), Bariloche, Argentina}

\author{Liliana Arrachea}
\affiliation{Departamento de F\'{\i}sica de la Materia Condensada,
and \\
Instituto de
Biocomputaci\'on y F{\'i}sica de Sistemas Complejos, Universidad de Zaragoza,
   Corona de Arag\'on 42, 50009 Zaragoza, Spain}

\author{Victor A. Gopar}
\affiliation{Instituto de Biocomputaci\'on y F{\'i}sica de Sistemas Complejos,
Universidad de Zaragoza,
   Corona de Arag\'on 42, 50009 Zaragoza, Spain}


\begin{abstract}
We calculate the distribution of the conductance $G$ in a one-dimensional  
disordered wire at
finite temperature $T$ and bias voltage $V$ in a independent-electron
picture and assuming full coherent 
transport. At high enough temperature and bias voltage, where several 
resonances of the system contribute to the 
conductance, the distribution $P(G(T,V))$ can be represented with good accuracy
by autoconvolutions of the distribution of the conductance at zero
temperature and zero bias voltage. The number of convolutions depends on $T$
and $V$. In the 
regime of very low $T$ and $V$, where only one resonance is relevant 
to $G(T,V)$, the conductance 
distribution is analyzed 
by a resonant tunneling conductance model. Strong effects of
finite $T$ and $V$ on the conductance distribution  are observed and 
well described by our theoretical analysis, as we verify by performing 
a number of 
numerical simulations of a  one-dimensional disordered wire at different
temperatures, voltages, and lengths of the wire. Analytical 
estimates for the first moments of $P(G(T,V))$ at high
temperature and bias voltage are also provided.

\end{abstract}

\maketitle

\section{\label{intro}Introduction}
The continuous progress in the fabrication of small electronic circuits 
 has kept active the theoretical study of quantum electronic transport. For 
example, transport properties in point contacts,
atomic chains, carbon nanotubes, and quantum wires are currently
 under experimental investigations  
\cite{{agrait},{nanot},{semi1d},{expefinV1},{expefinV2},{expefinV3}}.
Although the theoretical study of electronic transport in mesoscopic 
systems--where the phase 
coherence of electrons is preserved along the whole device--
has been of great interest for several decades, these recent experimental  
advances have further renewed
the motivation of theoreticians for studying the electronic 
transport  through  one-dimensional structures. 

In a disordered mesoscopic wire the random position of impurities gives a
stochastic character to the electronic transport. Therefore, it is of 
particular relevance the analysis of the statistical properties of
transport quantities like the conductance. On the theoretical side,
there is a full description 
of the statistical properties of the conductance in quasi-one-dimensional 
disordered systems within 
an independent-electron picture
at zero temperature $T$ and ``zero bias'' (actually, infinitesimally 
small voltage) $V$. \cite{pier_book,beenakker_review} 
This degree of detail in the theoretical description is, however, not available
for disordered wires at
finite temperatures and bias voltage, even in the simple case of 
one-dimensional (1D) wires. This is an unfortunate fact since experiments 
are usually performed within a wide 
range of $T$ and $V$. For example, the effect of finite 
bias voltage
has been found to modify the behavior of the conductance fluctuations.  
\cite{expefinV1,expefinV2,expefinV3} 

Several efforts have been made in order to incorporate the information of a
finite $T$ into the statistical description of quantum electronic transport 
in 1D-disordered systems. In Ref. \onlinecite{azbel}, at zero voltage, 
a model 
of zero-width resonances represented by $\delta$-functions was 
introduced in order to calculate the dependence on the temperature of the 
averages of the conductance $\langle G(T) \rangle$ and $\langle  \ln G(T) \rangle$. 
Assuming full coherent transport, 
in  Ref. \onlinecite{mosko}  the  effect of the
thermal smearing on the mean resistance of a 1D wire was studied.

At finite temperature and bias voltage, the distribution of the 
conductance $P(G(T,V))$  was calculated in Ref. \onlinecite{epl} by using 
a statistical model of resonant tunneling transmission. The methodology 
there employed was, however, 
restricted to disordered wires of length $L \gg \l$, where $l$ is the mean
free path,  and very small values of $T$ and $V$ (although sizable): typically
 $eV$ and $kT$ were considered  of the order or smaller than the 
mean spacing
between energy levels $\Delta$. Within this regime of temperatures and
voltages it can be assumed that only one
resonance--the 
closest to the Fermi energy-- contributes to the transport. Even with these 
limitations, that work shows that the 
conductance distribution do display novel features as a consequence of finite
$T$ and $V$ values. Particularly, it was shown  that the distribution of 
conductances narrows as the value of $T$ and/or $V$ is increased.

We would like to remark that the assumption of phase coherent transport
adopted in previous works, as well as in the present one, is
restricted by the phase relaxation  $L_\phi$ at finite temperature and
voltage. For example, inelastic scattering and thermally activated 
hopping processes,\cite{mott,stone,serota} which might be important at 
finite values of $T$, $V$ 
are not considered in our analysis. 
In Ref. \onlinecite{epl}, however, it 
is discussed the possibility of 
satisfying $ L_\phi > L$ even at finite temperature and bias voltage for 
short enough wires, which opens the possibility of observing the effect 
of finite
$T,V$ on the statistical properties of the conductance in realistic systems. 
It is also interesting to note that the range of voltage
$V$ where the averaged charge current $\langle J(T,V)\rangle$ is a linear
function of $V$ or departs slightly from this behavior, i.e., the 
so-called {\em linear  response} regime,  
can be rather wide as we will  show in Section \ref{results}.

The main goal of the present paper is to  compute the 
distribution of the
conductance $G(T,V)$ at values of $eV$ and $kT$ typically larger than
the mean level spacing, where several resonances
contribute to the conductance. We show that, within this regime of temperatures
and voltages, the distribution of the 
conductance can be  well described in terms of 
the convolutions of the zero temperature/voltage conductance  
distribution $p(g)$. 
The number of distributions to be convolved is determined by the 
number of resonances 
in an energy window where electron transport can take place. This energy window
is defined by the value of the temperature and voltage. 
The method introduced in the present work applies to
any degree of disorder i.e. {\em to any value} of the ratio $L/l$, provided
several resonances contribute to the transport,  allowing
for the investigation of short wires where coherent transport is
more likely to be observed. For completeness in
the sense that all regimes of temperature and voltage will be covered in this
paper, we also study the regime of small $T$ and $V$ following
Ref. \onlinecite{epl}

This paper is organized as follows. In Sec. \ref{method} we present the 
methodology
to analytically calculate the conductance distribution $P(G(T,V))$. 
We divide this section
into two subsections. The first one, \ref{resonant_model}, is devoted to
review the 
case of small 
temperatures and voltage regimes where one can assume that only the closest
resonance to the Fermi energy contribute to the conductance. \cite{epl}   
In the second one, subsection \ref{convolution_method}, we introduce a 
method based on the convolution of the known distribution of conductances 
at zero temperature and bias voltage in order to study the regime 
of high temperatures and bias voltages. 

In section \ref{results} we present the  model used in our
simulations and show some general features of our 1D-disordered system 
obtained numerically, which support the assumptions 
introduced in our theoretical proposal of  section \ref{method}. 
In section  \ref{results} we also compare the results of 
the distribution $P(G(T,V))$ from the
numerical simulations to the predictions of the theoretical approach 
presented in the previous section,
 for the different
regimes of $T$ and $V$. Finally, in 
section \ref{summary} we give a summary and conclusions of 
our study of the distribution of conductances at finite temperature and bias
voltage.

\section{Methodology}\label{method}
We consider the usual setup where the disordered conductor is placed between
left and right reservoirs at the same finite temperature $T$ and 
chemical potentials 
$\mu_L= \mu + eV/2$ and $\mu_R= \mu - eV/2$, respectively.
The conductance $G(T,V)=J(T,V)/eV$  
can be written as (in units of  the conductance quantum $2e^2/h$)
\begin{equation}
\label{gtv}
G(T,V)=\!\!\frac{1}{eV}\!\!\int_{-\infty}^
{\infty}\!\!\!\!\!\!\!\!\!dEg(E)\left(\!f(E-eV/2,T)- \!\!f(E+ eV/2,T)\right)\!,
\end{equation}
where $f(E,T) = \left\{\exp [(E-\mu)/kT] + 1\right\}^{-1}$ is the Fermi
function, being $k$ the Boltzmann constant, while $g(E)$ is the dimensionless
conductance for $E=\mu$ at $T=0$. It is clear from the above integral
expression for $G(T,V)$ that the energy window where
electronic transport takes place  will depend on the values of $T$ and $V$.
This can be emphasized recasting  the difference in the Fermi functions in the
above expression, Eq.(\ref{gtv}), as \cite{bagwell}
\begin{eqnarray}
\label{dff}
  f(E-eV/2, T)-f(E+eV/2,T) & = &    \nonumber \\  
 \left(\Theta (E-eV/2) -\!\Theta (E +eV/2) \right) \; 
&\ast&\frac{- \partial f(E,T)}{\partial E}  ,
\end{eqnarray}
where the symbol $\ast$ denotes the convolution in energy, $\Theta(x)$ is the 
unit step function, and 
$\partial f(E,T)/\partial E$  is the thermal smearing 
function. 
Therefore, the difference of the  Fermi functions can be expressed as a 
convolution in energy of two functions: one depends only 
on the applied voltage $V$ and the other one, only on temperature $T$.
This implies that thermal and voltages effects are statistically 
independent, and  this  will be
useful to calculate the distribution of $G(T,V)$.

\subsection{Small temperatures and bias voltages: one-resonance
contribution to $G(T,V)$ }\label{resonant_model}

In this subsection we briefly review the resonant model introduced in  
Ref. \onlinecite{epl}.   This model is appropriate for systems with non
overlapping resonances, which is satisfied for disordered systems with $L > l$.
 Assuming a Lorentzian line shape for 
the resonances, the dimensionless conductance $g(E)$ at $T=0, V\sim 0$ can be 
written as \cite{epl, mucciolo}
\begin{equation}
\label{g_model}
g(E)=\sum_\nu \frac{\Gamma_\nu^{(l)} \Gamma_\nu^{(r)}}
{(E-E_\nu)^2+{\Gamma_\nu^2}/4} ,
\end{equation}
where $\Gamma_\nu = \Gamma_\nu^{(\ell)} + \Gamma_\nu^{(r)} $ is the total 
width, 
$\Gamma_\nu^{(\ell, r)} \propto \Delta \exp [- (L\pm 2z_\nu)/\xi_\nu] $ are  
the left and right partial widths, and $\xi_\nu$ are the localization
lengths. 
On resonance $(E=E_\nu)$, the terms of the sum Eq. (\ref{g_model}),
$t_\nu=[\cosh(2z_\nu/\xi_\nu)]^{-2} $,
depend on the location $z_\nu$ of the state,
and are maximum for $z_\nu = 0$ (center of wire). Substituting Eq. 
(\ref{g_model}) into Eq. (\ref{gtv}), it is 
found that the conductance $G(T,V)$ is given by 
\begin{equation}
\label{G_energyintegrated}
G = - 2\pi kT {\rm Re}\left[ \sum_{\nu} \sum_{n=0}^\infty
\frac{t_\nu \Gamma_\nu}{(E_\nu - \mu + i\Gamma_\nu + i\omega_n)^2
 - \left( \frac{eV}{2}\right)^2} \right],
\end{equation}
where $w_n=(2n+1)\pi kT$. We now assume that $\Gamma_\nu << kT,eV 
{{<}\atop{\sim}} \Delta$. Under this condition only the resonance closest
to the Fermi level ($\nu=0$) contributes to the 
conductance $G(T,V)$ and therefore Eq. (\ref{g_model}) can be simplified to
\cite{epl} 
\begin{eqnarray}
\label{G_T_inter}
&&G(T,V) =
\frac{\pi}{2}\frac{t_0\Gamma_0}{eV} \nonumber \\ && 
\times\left[ \tanh \left( \frac{E_0 -\mu -
eV/2}{2kT} \right)
-  \tanh \left(  \frac{E_0- \mu + eV/2}{2kT} \right) \right]. \nonumber\\
\end{eqnarray}

In order to calculate the distribution of the
conductance $P(G(T,V))$ a statistical model for the 
resonances is introduced: the
resonances $E_0$ and their positions $z_0$ are considered uniformly
distributed in a interval $\Delta$ and $L$, respectively, while the 
distribution of the inverse of the localization 
lengths $p(x_0=2L/\xi_0)$ follows  a
Gaussian distribution with mean $\langle x_0 \rangle=L/l$ and
$\mathrm{var}(x_0)=2\langle x_0 \rangle$. This distribution for 
$p(x_0)$ is good for systems with  $l \ll L$. In Section \ref{results} 
we will show some examples for the 
distribution $P(G(T,V))$ obtained  within this resonant tunneling 
model.


\subsection{Large temperatures and bias voltages: contribution from 
several resonances to $G(T,V)$}\label{convolution_method}

We now go to the main goal of this work and propose a description 
of the conductance distribution, which is 
valid for arbitrary length of the wire and for  voltages and temperatures
satisfying $eV, kT > \Delta$, but small enough to satisfy that the behavior
of the average current does not depart from linear response.

Firstly, we discretize the integral in Eq. (\ref{gtv}) as 
\begin{eqnarray}\label{gdiscreteT}
G(T,V) & & \approx  \frac{\delta E}{eV} \; \sum_{i=1}^\infty
g(E_i)[f(E_i-(\mu-eV/2), T) \nonumber \\
& & -f(E_i- (\mu+eV/2),T) ] ,
\end{eqnarray}
where we have assumed the same width $\delta E$ for all the elements in  
the sum, Eq. (\ref{gdiscreteT}) . 

\subsubsection{Zero temperature, finite bias voltage}\label{subt0v}

Let us first consider the simple case of finite bias voltage and zero 
temperature. In this case, the thermal broadening function in Eq. (\ref{dff}) 
is a delta function 
and therefore  $g(E)$ in Eq. (\ref{gtv}) is only 
multiplied by a rectangular function of width $e V$. Thus 
Eq. (\ref{gdiscreteT}) is reduced to 
\begin{equation}\label{gdiscrete}
G(T=0,V) \approx  \frac{1}{N} \; \sum_{i=1}^N g(E_i) \; ,
\end{equation}
where the number of terms in the sum  $N$ satisfies $N \delta E $= $e V$. 
This approximation is exact  in the limit $N \rightarrow
\infty$. However, being our aim the evaluation of $P(G(T,V))$, we
approximate (\ref{gdiscrete}) by a finite number $N$  of {\em statistically
independent} contributions. We now assume that this number corresponds
to the mean number of levels in an energy
window $eV$, i.e., $N= eV/\Delta $. 
This is a  natural assumption since 
for a given disorder realization $g(E)$ is a spectral function with peaks at
the energy levels of the wire (resonances). 
In a non-interacting electron picture, the height of 
the resonance peaks, as well as
the energy levels change for  different disorder
realizations, but the average separation between levels is a well defined
quantity and the spectral weights centered at the different energy levels
are uncorrelated. In addition, we assume that $g(E)$ is a random stationary
function of the energy, at least in the energy window where transport takes
place. Then the statistical properties of $g(E)$ do not change in such
energy window and therefore the distributions $p(g(E_i))$ are actually
independent of the energy  $E_i$, i.e. $p(g(E_i)) \equiv p(g)$.  

Under the above assumptions the distribution $P(G(T=0,V))$ can be 
computed from  the convolution of $N$ distributions $p(g)$, i.e., the 
$N$th autoconvolution
of the distribution at zero temperature and bias voltage:
\begin{equation}
\label{pofgconvolutions}
P(G(T=\!0,V))=p^{(1)}(\hat{g})\ast p^{(2)}(\hat{g})\ast \cdots \!\!
\ast p^{(N)}(\hat{g})  ,
\end{equation}
where we have defined  $\hat{g} \equiv g /N$, and the upper indices enumerate
the number of distributions $p(g)$ that enters into the convolution.

As we mentioned in the Introduction, at zero temperature and infinitesimal
small bias voltage the
statistical properties of the transport quantities, in particular the
distribution $p(g)$ for  1D and quasi-1D 
disordered systems is  well 
known. In a 
framework of random-matrix
theory, a diffusion equation known as 
Dorokhov-Mello-Pereyra-Kumar (DMPK) equation has been
successful in describing the evolution of the conductance distribution 
$p(g)$ as a function of the system length $L$ in quasi-one
dimensional systems \cite{pier_book}. For strictly 1D wires the DMPK equation 
is reduced to 
the Melnikov equation whose solution $p(g)$ can be written 
as  \cite{beenakker_review}
\begin{equation}
\label{pofg}
p(g)=\frac{1}{\sqrt{2\pi}}\Big(\frac{1}{s}\Big)^{\frac{3}{2}}
\frac{{\rm e}^{-s/4}}{g^2}\int_{y_0}^{\infty}dy\frac{y{\rm e}^{-y^2/4s}}
{\sqrt{\cosh{y}+1-2/g}},
\end{equation}
where $y_0={\rm arccosh}{(2/g-1)}$ and $s=L/l$ is the length of the system $L$
in units of the mean free path $l$.
In the limit of $s > >  1$ and  $s << 1$ closed analytical expressions for
$p(g)$ can be obtained.\cite{{beenakker_review},{beenakker_rejaei}} We remark 
that the only parameter that enters in the distribution $p(g)$ is the 
so-called disorder parameter $s=L/l$.

The simple relationship between the random variables $G(T=0,V)$ and $g$,
Eq. (\ref{gdiscrete}), encloses important conclusions since this
 means that the Cumulant  Generating function for $G$ and $\hat g$ are 
also simple related:  
\begin{equation}
\log {M_{G}(\Lambda)} = N \log{(M_{\hat{g}} (\Lambda))} \; ,
\end{equation}
 with  $M_{\gamma}(\Lambda) \equiv \int \exp({-\Lambda \gamma} ) \; P(\gamma)
 d\gamma$ being $\gamma
=G, \hat{g}$, while for the average and the first three central moments 
$\Sigma_q \equiv \left\langle ( G - \left\langle G \right\rangle )^{q} 
\right\rangle$ and 
$\sigma_q = \left\langle  ( g - \left\langle g \right\rangle )^{q} 
\right\rangle $ it is satisfied: 
\begin{eqnarray} \label{cumu}
\left\langle G \right\rangle & = &\left\langle g \right\rangle  \; , \\
 \Sigma_q & = & \frac{1}{N^{(q-1)}} \sigma_q \ \ \ \ \ \ \mathrm{for}  \ \  
q=2,3 \;  \label{cumu2}. 
\end{eqnarray}
$\sigma_1 = 0$ by definition.
Then, Eq.(\ref{cumu}) shows that the mean value of the 
conductance  $\left\langle G \right\rangle$  is independent of the 
bias voltage $V$. 
\begin{figure}
\includegraphics[width=0.8\columnwidth]{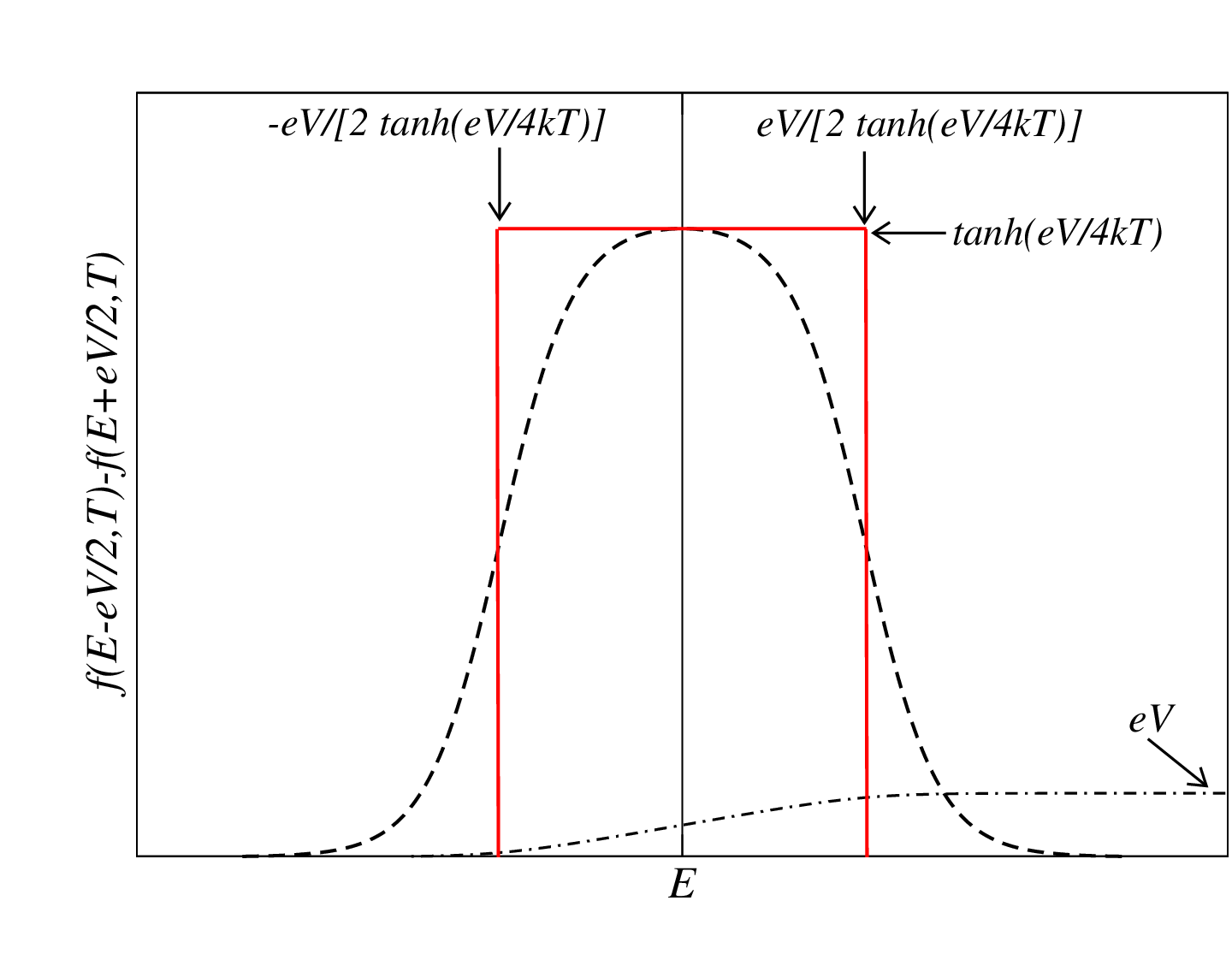}
\caption{\label{fig_sketch} (Color online) Sketch of the behavior of the 
function (\ref{finiT}) and its approximation by a rectangular 
function, in black dashed and red  solid lines. The behavior of 
the integral of the
  function (\ref{finiT}) is also indicated in dashed-dotted lines.} 
\end{figure}

\subsubsection{Finite temperature and bias voltage}\label{finitetandv}
Let us now  consider a more realistic situation where temperature
and bias voltage are both finite. 

At finite temperature, the difference of 
Fermi functions in Eq. (\ref{gtv}) reads
\begin{eqnarray}
\label{finiT}
f\left(E- (\mu-eV/2), T\right)&-&f \left(E- (\mu+eV/2),T \right) = \nonumber \\
 \frac{1}{2} \biggr[ \tanh \left(\frac{E- (\mu-eV/2)}{2kT} \right) &-& \tanh 
\left(\frac{E- (\mu + eV/2)}{2kT} \right) \biggr]. \nonumber\\
\end{eqnarray}
In fig. \ref{fig_sketch} it is shown the behavior of 
Eq. (\ref{finiT}). In order
to keep our 
statistical analysis of the conductance as simple as possible, we approximate the bell-shaped 
function (\ref{finiT}) by a rectangular function of height 
of $\tanh(eV/4kT)$ and width $eV/\tanh(eV/4kT)$. The area of the rectangle is, 
therefore, $eV$ as the area of the original bell-shaped function 
Eq. (\ref{finiT}). This
simplification allows us to proceed as in subsection \ref{subt0v}:
from the width $eV/\tanh(eV/4kT)$ we define an ``effective number of 
resonances'', $N_{eff}$, given by 
\begin{equation}\label{neff}
N_{eff} = \frac{eV}{\Delta \tanh(eV/4kT)} .
\end{equation}
We can verify that for $T \to 0$, $N_{eff} \to N=eV/\Delta$, as expected.
Thus to calculate $P(G(T,V))$ we can follow exactly the same procedure of the
previous subsection \ref{subt0v}, we just replace $N$ by $N_{eff}$. 
Also, at finite $T$ and $V$ it is  possible to define a simple 
relation between  
cumulants and moments of the distribution $P(G(T,V))$ and $p(g)$;  
again, substituting  $N$ by $N_{eff}$ in Eq. (\ref{cumu2}).


\section{Numerical results}\label{results}
In this section we verify our theoretical study of the statistical properties
of the conductance, in particular the conductance distribution, 
by comparing to numerical simulations of a 
1D-disordered wire.  The results presented here give numerical evidence 
that supports the main hypothesis of our theoretical model and 
benchmark the quality of the approximation of  the distribution function at
finite bias and temperature on the basis of convolutions of the distribution
in Eq.  (\ref{pofg}) against exact numerical results.

In our simulations we model a 1D-disordered wire of length $L$  using the 

standard tight binding
 Hamiltonian of spinless electrons  with a single atomic orbital   
per lattice site and  nearest neighbors hopping parameter $t$:
\begin{equation}\label{ham}
H^{wire}= - t \sum_{j=1}^{N_s} (c^{\dagger}_j c_{j+1} + H. c. ) + \sum_{j=1}^{N_s}
 \varepsilon_j c^{\dagger}_j c_j,  
\end{equation}
with $L=N_s a$, being $a$ the lattice constant, which we set as the unit of
length and $N_s$, the number of sites of the 1D  lattice. 
The on site energies $\varepsilon_j$ are chosen randomly 
from  an  uniform distribution of width $W$ (Anderson model).  
All the energy scales are taken dimensionless  in units  of the hopping
parameter $t$. The disordered conductor  is connected  to the left and to the
right to 1D clean semi-infinite leads, which we assume to be at 
chemical potentials $\mu_L= \mu + eV/2$ and  $\mu_R= \mu - eV/2$,
respectively, and temperature $T$.
The Hamiltonian describing the contacts between the reservoirs and the 
disordered wire reads:
\begin{equation}\label{hamc}
H^{cont}= - t (c^{\dagger}_1 c_{k_L} + c^{\dagger}_{N_s} c_{k_R} + H. c. ),
\end{equation}
being $k_{L,R}$ the contact sites of the left and right semi-infinite 
leads, respectively.

For each disorder realization, the current is numerically calculated 
from  the expression
(\ref{gtv}), with 
\begin{equation}
g(E)= \Gamma^2(E) |G^R_{1 N_s}(E)|^2,
\end{equation}
being $\Gamma(E)=  |t|^2 \Theta(|E|- 4t) \sqrt{16 t^2- E^2}/4t^2$, which
corresponds to reservoirs modeled by semi-infinite tight-binding chains with
hopping $t$, 
while $ G^R_{1N}(E) $ is the retarded Green's function for the disordered
chain connected to the reservoirs. \cite{datta} Typically, we generate
numerically 
$10^4$ to $10^5$ realizations of disorder. For several
values of the disorder parameter $s=L/l$, we have 
verified that for very small bias ($eV << \Delta$) and $T=0$ we are able to
reproduce the distribution $p(g)$ given by Eq. (\ref{pofg}). The mean free
path $l$ is extracted from the numerical simulations through the relation 
$\langle \ln g \rangle \equiv -L/l$. In all numerical 
simulations we present below we have fixed $\mu=0$ and the disorder strength to $W=0.5$ 
which sets $l=144$.

The theoretical distributions $P(G(T,V))$, as described by
Eq. (\ref{pofgconvolutions}) of the previous section, are obtained 
by collecting the data for $G(T,V)$ from 
an ensemble of conductances
generated numerically using a Metropolis Monte Carlo algorithm.

\subsection{General features}

\begin{figure}
\includegraphics[width=\columnwidth]{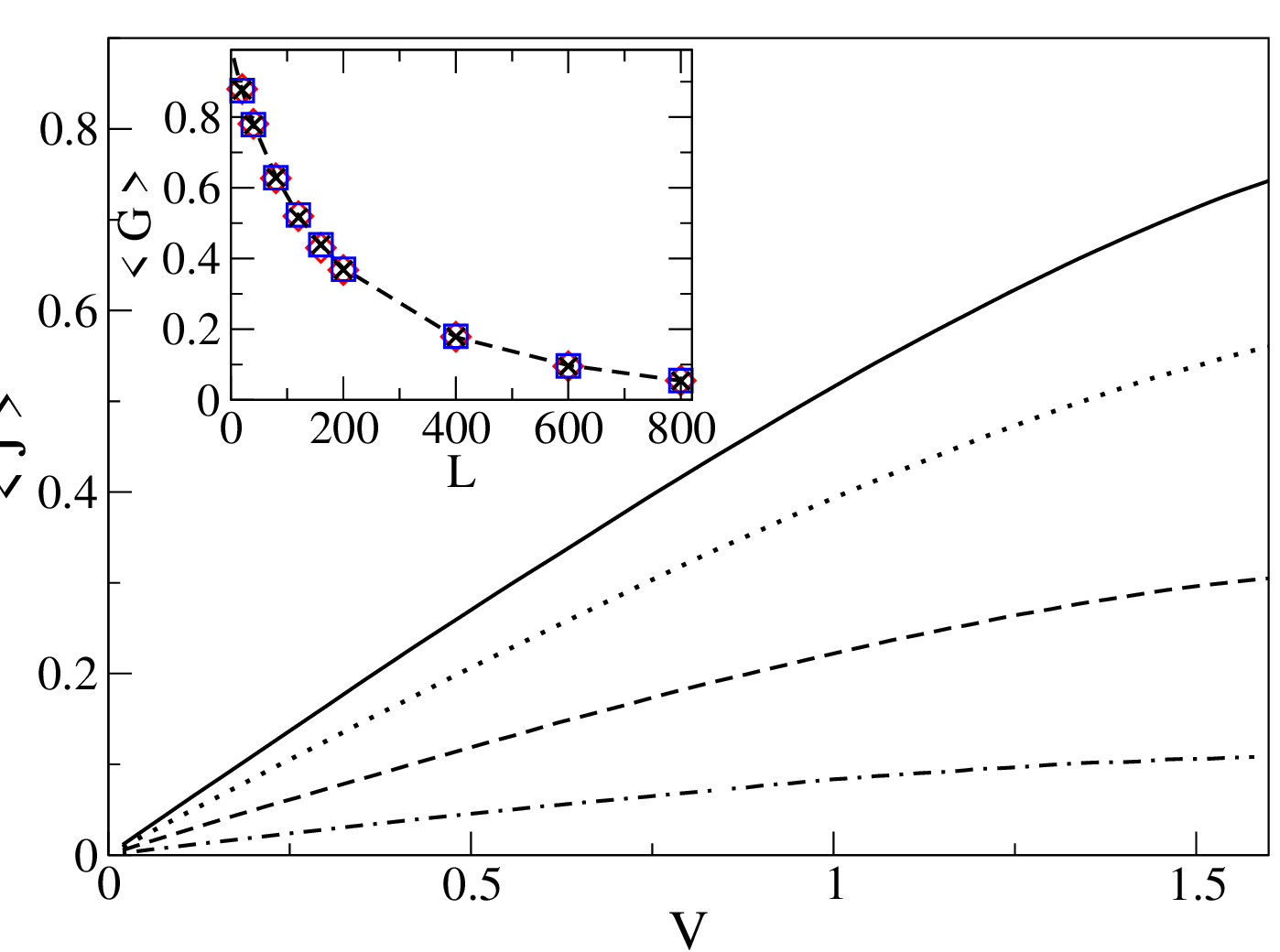}
\caption{ (Color online) \label{fig_J} $\langle J(T,V) \rangle$ as a 
function of $V$ for 
$W=0.5$, $\mu=0$ and different lengths 
$L= 120$ (solid ), $L=200$ (dotted), $L=400$ (dashed), and $L=800$ 
(dashed-dotted). Inset: $\langle G(T,V) \rangle$ vs $L$ for
$T=0$ and  $V= 0.01$ (black crosses), $V= 0.1$ (blue squares) and $V= 0.2$ 
(red diamonds). Note that the different  symbols  are  overlapped.}
\end{figure}

Before presenting our results for the probability distribution at finite
temperature and bias voltage, we would
like to establish the range of
the bias voltage $V$ where the
averaged current $\langle J(T=0,V) \rangle$  varies linearly with $V$. To
this end we have included in Eq. (\ref{ham}) the potential drop due to the
presence of the bias by modifying the local energies as $\varepsilon_j
\rightarrow \varepsilon_j + e V/2 - j e V/N_s $. 
In Fig. \ref{fig_J} we show the disorder averaged 
current $\langle J(T,V) \rangle$ as a function of the bias voltage,  
at $T=0$. A linear behavior of 
$\langle J(V) \rangle$   with the bias $V$ is observed up to
$V \lesssim 0.5$ for different lengths of the system.  Therefore 
if we want to restrict our study to the linear
response regime,  we cannot take arbitrary large
values of $T$ and $V$. In the present model, this implies 
 energy values for $eV$ and $kT{{<}\atop{\sim}}0.5$. Let us note, that 
this
 range of energy is actually rather wide: it is larger than 10$\%$ of the
 total band width of the clean system (equal to $4t$). Since we focus in
 voltages within the linear response regime, in what follows we consider 
the wire model (Eq. (\ref{ham})) without including the effect of the 
potential drop in the local energies $\varepsilon_j$.
 
In the inset of Fig. \ref{fig_J} we also show the average of 
the conductance $\langle G(T,V) \rangle$ for different 
values of $T$, $V$, as a function of the system length $L$. We observe 
that $\langle G(T,V) \rangle$ 
is independent of the values of the temperature and voltage
and decreases exponentially with $L$ as it is expected. This behavior 
is in agreement
with previous results \cite{epl} as well as with the prediction 
based on Eq. (\ref{cumu}). 

\begin{figure}
\includegraphics[width=\columnwidth]{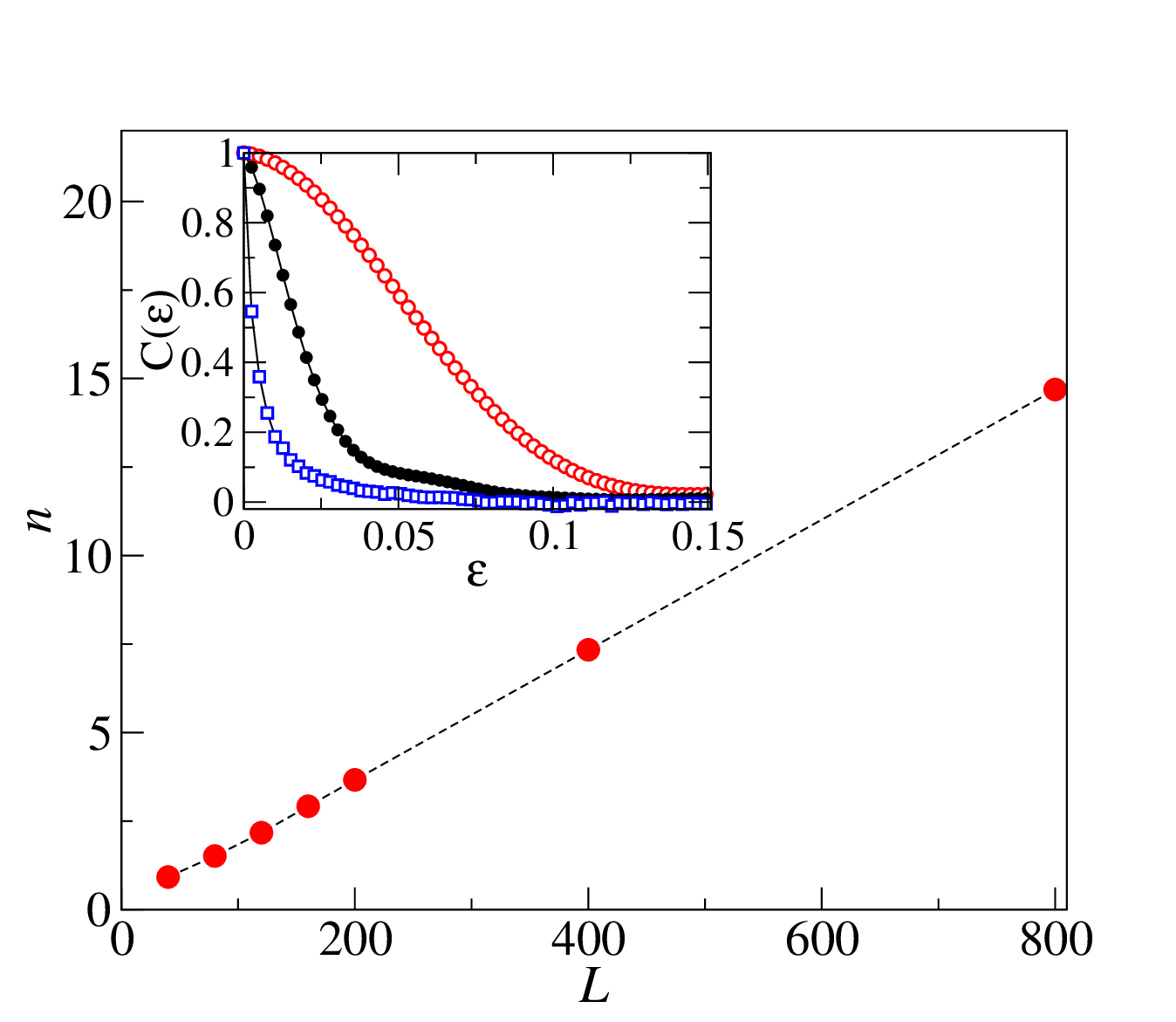}
\caption{\label{fig_levels} (Color online) Mean 
number of levels $n$ of the Hamiltonian (\ref{ham}) in an energy interval 
equal to $0.1$, as a function 
of the system length $L$. Inset: The autocorrelation function 
$C(\varepsilon)$ is 
plotted for $L= 40,\; (\Delta=0.108)$, $L=120,\; (\Delta=0.046)$ and $L=800,\;
(\Delta=0.007)$ 
(red open circles, black solid 
circles and blue
square symbols, respectively). See the text for details.}
\end{figure}

Another important point to verify is the reliability of our assumption that
the mean energy spacing between levels of Eq.(\ref{ham}) sets the energy scale
where the spectral weights of $g(E)$ are statistically independent. To this
end, we have investigated the behavior  of the
auto-correlation function $C( \varepsilon) \equiv 
\overline{\langle g(E)g(E + \varepsilon) \rangle-\langle g(E) 
\rangle \langle g(E+ \varepsilon) \rangle} $, where the over line means energy
average.   In Fig. \ref{fig_levels} we have plotted 
$C(\varepsilon)$ (normalized 
by dividing by $C(0)$) for three different 
wire lengths: $L= 40, 120$ and $800$. 
In all the cases $C(\varepsilon)$ decays between  $90\% \sim 99 \%$ of its  
value at $\varepsilon =0$
for  $\varepsilon  < \Delta $. In order words, $C(\varepsilon)$ is a vanishing
function in an energy scale $\varepsilon$ 
smaller  than the mean level spacing. 

We have also verified numerically that in the linear response regime 
$p(g(E))$ is in fact independent of $E$, as we assumed in the previous
section.

\begin{figure}
\includegraphics[width=\columnwidth]{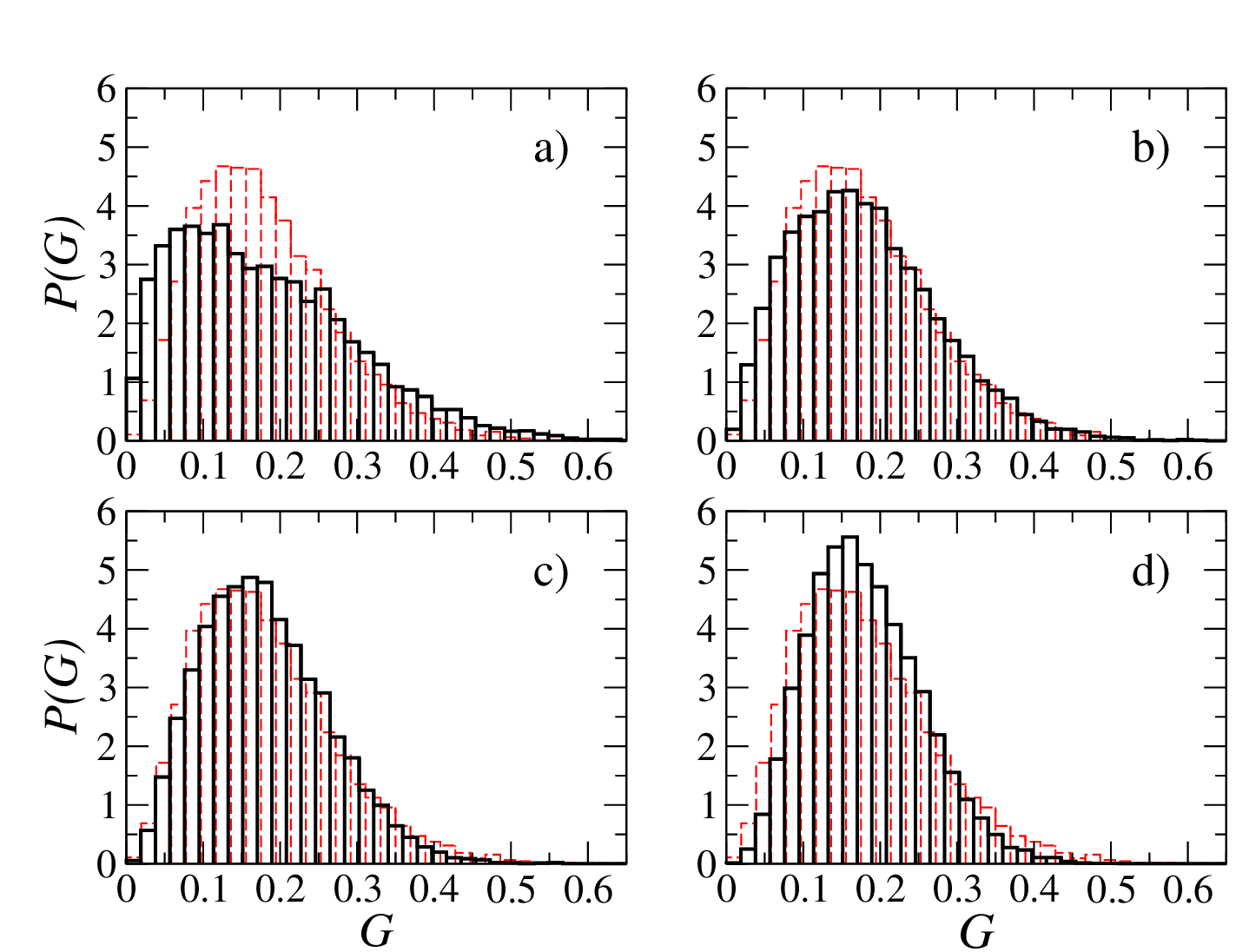}
\caption{\label{fig_illustrative} (Color online) Histograms  obtained  from 
Eq.(\ref{pofgconvolutions}) 
performing different number of convolutions $m$,
for a finite voltage $V=0.1$ and $L=400$.
a) $m=4$, b) $m = 6$, c) $m=8$ and d) $m=10$. For comparison the distribution  
$P(G)$ ( histogram in red dashed line) obtained  numerically is also 
plotted. For $m=8 \ \ (=N)$ 
convolutions the distribution $P(G)$ is quite well reproduced. The size bins
of the numerical and theoretical histograms are slightly different to 
distinguish better between both histograms.}
\end{figure}

On the other hand, it is instructive to show the evolution of the 
conductance distribution 
$P(G(T,V))$ with the number of autoconvolutions of $p(g)$, as described in the
 previous section. We have chosen the simple case of $T=0$ with 
and finite $V=0.1$ with $s=400/144 \approx 2.7$. In 
Fig. \ref{fig_illustrative}, the histogram in dashed
line corresponds to the distribution obtained from our numerical 
simulations, while
the histograms in solid line correspond to $P(G)$ obtained by $m$ 
autoconvolutions of $p(g)$, for $m=4,6,8$, and 10. We can observe that as $m$
increases from 4 to 6 the theoretical $P(G)$ evolves to the numerical
distribution. When $m=N=8$, which
corresponds to the mean number of levels $N$ for $L=400$, see Fig. 
\ref{fig_levels}, a very good agreement is found with the numerical
distribution. The fact that the optimum number of autoconvolutions of $p(g)$
coincides at $T=0$ with the mean number of levels within eV,
supports the validity of our approach of 
subsection \ref{subt0v}. 

\subsection{Small temperatures and bias voltage: $P(G(T,V))$ from one 
resonance contribution to $G(T,V)$} 

In this subsection we show some numerical results for the 
distribution $P(G)$ when the values of 
the temperature and bias voltage are small  enough  that only one resonance
contributes to the conductance $G(T,V)$ (subsection \ref{resonant_model}). 
We also indicate the limitations of this method by showing an example of the
conductance  distribution at regimes of $T$ and $V$ beyond the scope of 
the resonant model.

\begin{figure}
\includegraphics[width=\columnwidth]{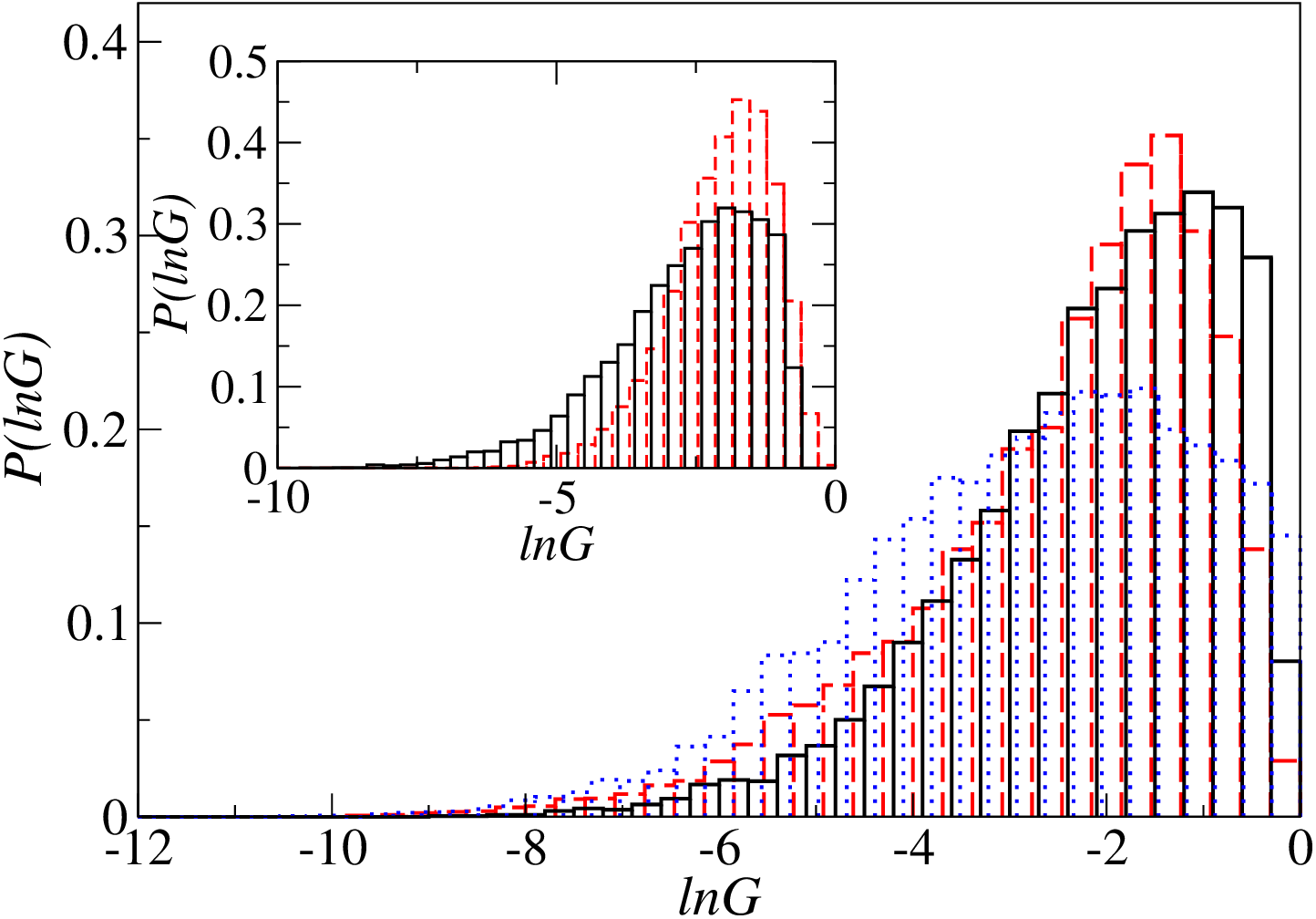}
\caption{\label{reso_fig} (Color online)
 Histograms  obtained with the resonant 
model of subsection \ref{resonant_model} (solid black lines) for $L=400$. 
Results in dashed red lines correspond to the numerical simulations on
the disordered tight-binding chain. Plots in
the main panel correspond 
to $V=\Delta, T=0 $ while the inset corresponds to $V=\Delta, T=\Delta/3$.
The distribution $p(g)$ is also shown in the main panel (blue dotted line).
}
\end{figure}
The theoretical distribution $P(G(T,V))$ is calculated by 
generating numerically an ensemble of conductances $G$ 
accordingly to Eq. (\ref{G_T_inter}), where  the  random variables  
$E_0$ and $z_0$ are obtained numerically from a uniform distribution, while
$x_0$ is extracted from a Gaussian distribution, as it is described in 
\ref{resonant_model}.
 
In figure \ref{reso_fig} we show the distribution of $\ln G$ (we have chosen
the variable $\ln G$ since the details of the distribution can be better seen
in the logarithm scale) from the resonant tunneling model (histogram in solid
line) for the case of $L=400 \sim 3 l$ and $e V = \Delta$. A good agreement
with the numerical simulation (histogram in dashed line) is seen. 
Note that,
although this situation corresponds to a small bias voltage, the exact
distribution (solid lines) 
clearly departs from the behavior of the ``zero'' bias distribution 
$p(g)$ (histogram in dotted lines).
Therefore the resonant model reasonably describes the numerical  
behavior of $P(G)$, in particular 
the decrease of the width of the distribution as $V, T$ increases (with $V, T
< \Delta$). However, if we increase the value of $T$, keeping 
$eV=\Delta$, in our previous numerical example, the 
quality of the approximation
provided by this model deteriorates, as it is illustrated in the inset of the
figure \ref{reso_fig}. This happens in general when the values $T$ 
and/or $V$ are increased 
in such a way that
more than one resonance might contribute to the conductance. In the next
subsection we will see that the convolution method describe correctly 
the regime of $T$ and $V$ when several resonances are involved in the 
transport problem. 

\subsection{Large temperature and bias voltage: $P(G(T,V))$ from 
several resonances contributions to $G(T,V)$}

\subsubsection{Zero temperature and finite bias voltage}

\begin{figure}
\includegraphics[width=\columnwidth]{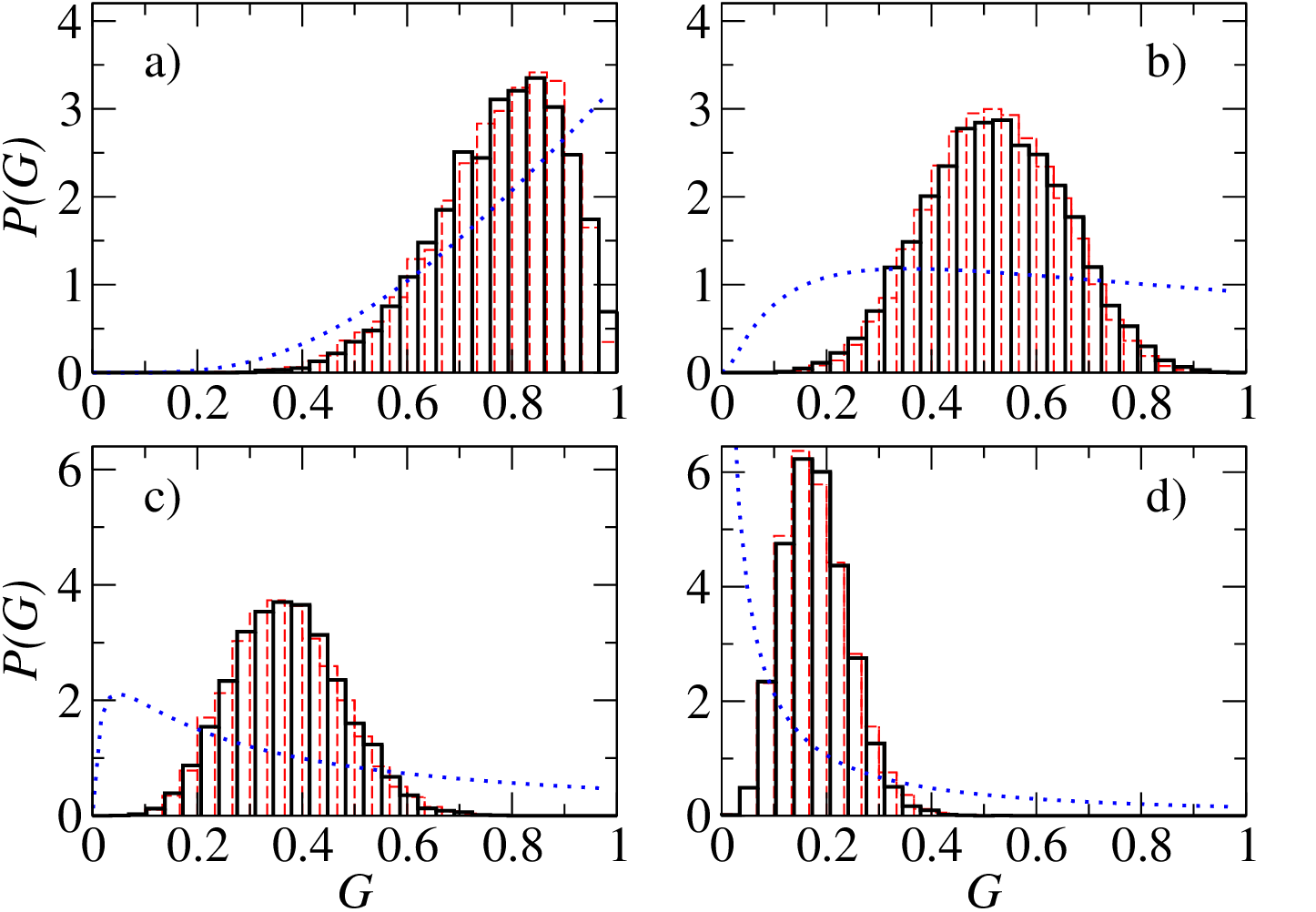}
\caption{\label{fig_VTzero} (Color online) Distributions $P(G)$ for  
$V=0.2$ and $T=0$ 
(red dashed line histogram) obtained numerically for different lengths 
a) $L=40$, b) $L=120$, c) $L=200$, and d) $L=400$. The solid black line 
histograms correspond to the distribution of $G$ obtained as a convolution 
of $N$ terms a) $N=2$, b)$N=4$, c)$N=7$, d)$N=14$. The blue dotted curves show 
$p(g)$ for zero bias and temperature. The size of the bins
of the numerical and theoretical histograms are slightly different to 
distinguish better between both histograms.} 
\end{figure}

Let us now go to the analysis of the probability
distribution function at finite $V$ with $T=0$ using the methodology introduced
in subsection \ref{subt0v}.
In figure \ref{fig_VTzero} we show the distributions $P(G(T=0,V))$
for different lengths $L$  of
the  disordered wire with $V=0.2$. We recall that $l=144$ in all
cases.  For each length $L$, the distribution $P(G)$ obtained  from
Eq.(\ref{pofgconvolutions}) (histogram in solid line) and the numerical
distribution (histogram in dashed line) are both displayed for their 
comparison. A very good agreement can be seen. 
In order to provide evidence of the strong effect 
of the finite bias voltage on the conductance distribution, 
$p(g)$ is also shown in dotted lines in the same figure.

We have also verified the relation between the second and third moments,
Eq. (\ref{cumu2}). For a bias voltage $V=0.2$, in Table \ref{table1}  we 
show the values of $\Sigma_{q}$ extracted  from the numerical 
simulation and those for $\sigma_{q}$ ($T=V=0$) calculated from then integral
expression of $p(g)$,  Eq. (\ref{pofg}). Again,  a
good agreement between numerics and theory is obtained.
\begin{table}
\caption{moments $\Sigma_q$ ($q= 2, 3$) for bias voltage
  $V=0.2$, $W=0.5$ for different lengths $L$ obtained  
numerically and $\sigma_q$ from the distribution $p(g)$, Eq.(\ref{pofg}).}
\begin{tabular}{|c|c|c|c|c|c|}  
    \hline 
    \;$L$ \;&\; $N$\; & \; $\Sigma_{2}$  \;& \;$\sigma_{2}/N$ \;&    \;$\Sigma_{3}$ \;& \;$\sigma_{3}/N^2 $ \; \\ 
    \;  \;&\; \; & \; $\times 10^{-2}$ \;& \; $\times 10^{-2}$ \;&    \;$
    \times 10^{-3}$ \;& \;$\times 10^{-3} $ \; \\ 
    \hline \hline
    \;40 \;& \; 2 \; &\; 1.469 \;& \; 1.427 \;&\; -1.32 \;&\; -1.01 \; \\ 
    \hline 
    \;80 \;& \; 3 \;& \;1.716 \;& \; 1.802 \;&\; -0.574 \;&\; -0.460 \; \\ 
    \hline 
   \;120 \;&\; 4 \;&\; 1.554 \;& \; 1.733 \;&\; 0.010  \;&\; 0.054 \; \\
    \hline
    \;200 \; & \; 7 \;&\; 1.099 \;&\; 0.933 \;&\; 0.32 \;&\; 0.263 \; \\
    \hline
   \;400 \; & \; 14 \; & \; 0.414 \;&\; 0.383 \;&\; 0.160 \;&\; 0.110 \; \\
    \hline
    \;800 \;& \; 29  \;& \; 0.065 \; & \; 0.071 \;&\; 0.015 \;&\; 0.012 \; \\
    \hline
\end{tabular}
\label{table1}
\end{table}

\subsubsection{Finite temperature and bias voltage}
\begin{figure}
\includegraphics[width=\columnwidth]{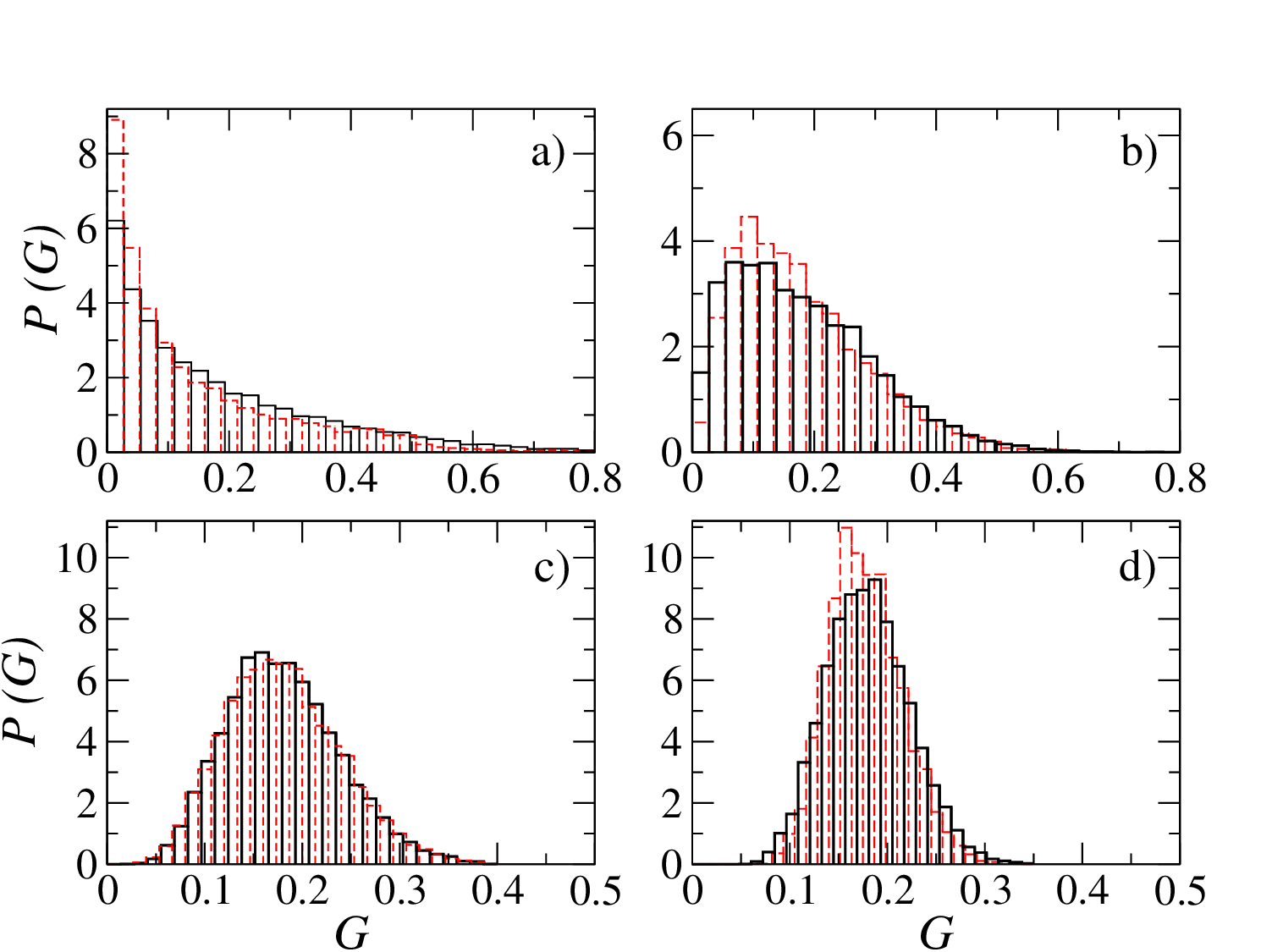}
\caption{\label{fig_finitetv} (Color online) Numerical distributions $P(G)$
  (dashed red line
 histograms) of a system with 
length $L=400$, $V=0.01$, and different temperatures: a)$T=0.005$, 
b)$T=0.01$, c)$T=0.05$, d)$T=0.1$. The theoretical
 distribution from $N_{eff}$ autoconvolutions of $p(g)$ (solid line
 histograms) are compared with the corresponding numerical results: 
a)$N_{eff}=2$, b)$N_{eff}=3$, c)$N_{eff}=15$, d)$N_{eff}=29$. A good agreement
 is observed.}
\end{figure}

Finally, let us consider both finite temperature and bias voltage. This
case corresponds to the regime considered in subsection \ref{finitetandv}. There, an 
effective number of resonances was introduced, Eq. (\ref{neff}), which
corresponds  to 
the number of autoconvolutions of the distribution $p(g)$, Eq. (\ref{pofg}), to
be used in order to obtain $P(G(T,V))$. In Fig. \ref{fig_finitetv} we 
present the results for $P(G(T,V))$
(histograms in solid line) from the convolution method  for  $L=400$, 
$V=0.01$, and 
four different   values of $T$. For each value of $T$, the 
number of 
convolutions $N_{eff}$ changes according to Eq. (\ref{neff}). As in previous
cases, we compare to the numerical simulations 
(histograms in dashed line). A good agreement is seen in all cases.

The first moments of the  conductance distribution were also obtained. The
results are shown in Table \ref{table2}, where, as 
before,  $\Sigma_{q}$ is  obtained  from the numerical
simulation, while $\sigma_{q}$ is computed from  the expression for 
$p(g)$, Eq. (\ref{pofg}). The agreement 
between the numerical
simulation and theory 
is, in general, reasonably good. We recall that at finite $T$ we have 
introduced 
the simplification of  representing
the difference of the Fermi functions by a rectangular function, Eq. 
(\ref{finiT}). This additional approximation might be the cause for the discrepancies
between $\Sigma_q$ and $\sigma_q/N_{eff}^{q-1}$ observed for 
small values of $N_{eff}$  (first two lines of table
II). Notice that a small error in the estimate of the integer number  
$N_{eff}$ 
implies a large relative error for small $N_{eff}$. In all cases, however,
 the behavior of the moments and the conductance distribution
is well described by our method.

\begin{table}
\caption{Moments $\Sigma_q$ ($q= 2, 3$) for bias voltage
  $V=0.01$, $W=0.5$ and several temperatures $T$ for  $L=400$ obtained  
numerically and $\sigma_q$ from the distribution $p(g)$, Eq.(\ref{pofg}) .}
\begin{tabular}{|c|c|c|c|c|c|}
    \hline
    \;$T$\;&\;$N_{eff}$\;&\;$\Sigma_{2}$\;&\;$\sigma_{2}/N_{eff}$\;& \;
$\Sigma_{3}$\;&\;$\sigma_{3}/N_{eff}^2$ \; \\
   \;\;&\; \;&\;$\times 10^{-2}$\;&\;$\times 10^{-2} $\;& \;$\times 10^{-3}$\;&\;$\times 10^{-3}$ \; \\
    \hline \hline
    \;0.005 \;& \; 2 \;& \;1.964 \;& \; 2.585 \;&\; 3.24 \;&\; 5.0 \; \\
    \hline
    \;0.01 \;& \; 3 \;& \;1.175 \;& \; 1.723 \;&\; 1.239 \;&\; 2.2 \; \\
    \hline
   \;0.05 \;&\; 15 \;&\;0.294 \;& \; 0.344 \;&\; 0.107 \;&\; 0.088 \; \\
    \hline
    \;0.1 \;& \; 29 \;& \;0.144 \;&\; 0.178 \;&\; 0.023 \;&\; 0.023 \; \\
    \hline
\end{tabular}
\label{table2}
\end{table}

Finally, it worth mentioning that we have verified that the approach based in
convolutions of $p(g)$ also provides a good approximation to the exact
$P(G(T,V))$ at finite $T$ in cases with larger $eV$ (e.g. $eV=0.1,0.2$).

\section{Summary and conclusions}\label{summary}

In an independent electron picture and assuming full phase-coherent electronic 
transport, we have studied the distribution 
of the conductance $P(G(T,V))$ in 1D-disordered systems at finite 
temperature and bias voltage. 

We have observed a strong effect of finite $T$ and $V$ at the level of the
conductance 
distribution. In general $P(G(T,V))$ is narrower compared to the conductance 
distribution
at $T=0, V \sim 0$. The average of the conductance is, however, 
independent of $T$ and 
$V$ (see Fig. \ref{fig_J}). This implies that higher moments have to 
be analyzed to see the effect of the temperature and voltage.

When temperatures and voltages are small (less than the mean-level spacing),
only one resonance is relevant to the conductance. In this regime, the
distribution of the conductance is well described by a simplified 
resonant-tunneling model, Eq. (\ref{G_T_inter}), as we have verified
numerically.  
As the temperature and bias voltage is increased several resonances
contribute to $G(T,V)$. In this regime, 
the conductance distribution
$P(G(T,V))$ can be  obtained  from the 
convolutions of the known distribution of conductance at zero temperature 
and bias voltage $p(g)$. The number of autoconvolutions of $p(g)$ is 
determined  by the  
width of the energy window where transport can take place. In the case of
zero temperature and finite bias voltage, the width of the energy window is
trivially $eV$ and the number of convolutions is given by the mean number of
levels $N=eV/\Delta$. In the case of finite temperature and a finite 
bias voltage, 
we have simplified the problem by introducing an effective number of
resonances  that allow us to reduce the problem of finite $T$ and $V$ to the
simpler case of zero temperature, and finite $V$.  
The results of our theoretical method have been compared to 
numerical simulations of a 
1D-disordered system at different regimes of temperature, voltage, and
different values of the length
of the system. A good agreement has been found in all cases. 
We point out that for small $T$ and $V$ the line shape of the resonances was
relevant in the calculation of $G(T,V)$ (see Section \ref{resonant_model}).
When the values of $T$ and $V$ are such that several resonances contribute 
to the conductance, the line shape is seen to be irrelevant. 
To conclude, we
remark that with the resonant model and the convolution method of sections 
\ref{resonant_model} and \ref{convolution_method}, respectively,
we are able to analyze the conductance distribution at  
all regimes of temperature and bias voltage, under the assumptions described in
the paper.

\section{acknowledgments}

F. F is grateful to the hospitality of BIFI in Zaragoza and
CAB in Bariloche where part of this work has been done, and to AUIP for a
travel grant. M.J.S.  and L.A. acknowledge financial support
from  PICT 0311609 and (M.J.S.) 
from Fundaci\'on Antorchas 
 and PICT 0313829 from ANPCyT, Argentina. 
L. A. and V. A. G. also acknowledge financial support from the Ministerio 
de Educaci\'on y Ciencia, Spain, through the Ram\'on y Cajal Program. 
This work was supported by grant ``Grupo de investigaci\'on de excelencia DGA''
and  (L.A.) BFM2003-08532-C02-01 from MCEyC of Spain. L.A and  M.J.S. are
staff members and F.F is fellow of CONICET, Argentina.

\end{document}